\documentclass[a4paper,11pt]{article}
\usepackage{pos}
\usepackage[abs]{overpic}

\title{Prospects for measuring exclusive diffractive $\eta,\eta'$ at the LHC}

\author*[a]{Rainer Schicker}

\affiliation[a]{Phys. Institute, Heidelberg University\\
  Im Neuenheimer Feld 226, 69120 Heidelberg, Germany}

\emailAdd{schicker@phys.uni-heidelberg.de}

\abstract{Central exclusive diffractive production in proton-proton collisions
  at hadron colliders is characterised by hadronic activity at or close to
  midrapidity, and by the two forward scattered protons, or their remnants.
  In such events, no particles are produced between the midrapidity system
  and the forward beam particles. These events can hence be identified with
  appropriately placed detectors for measuring the forward scattered protons,
  or their remnants, and a detector system covering the midrapidity range.
  At the energies of the LHC, central diffractive production in proton-proton
  collisions is dominated by pomeron-pomeron fusion. The description of the
  pomeron within the Regge approach is summarized, and the feasibility of
  identifying pseudoscalar mesons  $\eta,\eta'$ in pomeron-pomeron fusion
  is studied for determining the spin structure of the pomeron.}

\FullConference{Proceedings of the Corfu Summer Institute 2025 "School and Workshops on Elementary Particle Physics and Gravity" (CORFU2025)\\
21 - 26 September, 2025\\
Corfu, Greece\\}

\begin{document}
\maketitle

\section{Diffractive event topologies}

In the Regge approach, the hadronic interaction between strongly interacting
particles is due to the exchange of trajectories. Such trajectories
$\alpha (t)$ parameterise the almost linear dependence between the spin and
the mass squared ($t  =  m^{2}$) of a particle and its higher spin excitations.
A trajectory $\alpha(t) = 1 + \varepsilon +  \alpha' t$ is characterised by
two parameters, the intercept $1  +  \varepsilon$ and the slope $\alpha'$.
The contribution of a trajectory to the energy behaviour of the total cross
section scales as $\sigma_{tot} \propto s^{\varepsilon}$. For meson trajectories,
such as the $\rho,\omega,a,f$, the value of $\varepsilon\!\sim\!-0.5$, and the
corresponding contributions hence decrease as function of energy \cite{PomQCD}.
Such a behaviour is seen in the elastic and total hadron-hadron cross section
which decrease from threshold up to a centre-of-mass energy
\mbox{$\sqrt{s} \sim$ 20 GeV.} At higher energies, however, elastic and total
hadron-hadron cross sections increase \cite{PDG}. Within the Regge approach,
this increase is attributed to the existence of an additional exchange
mechanism in both elastic and inelastic channels, which is refered to as
the pomeron trajectory. The characterisation of the bound states underlying
the pomeron trajectory, and their possible experimental verification,
is a considerable challenge from the theoretical as well as the
experimental perspective.

\begin{figure}[h]
\begin{center}
  \vspace{-.2cm}
  \includegraphics[width=.27\textwidth]{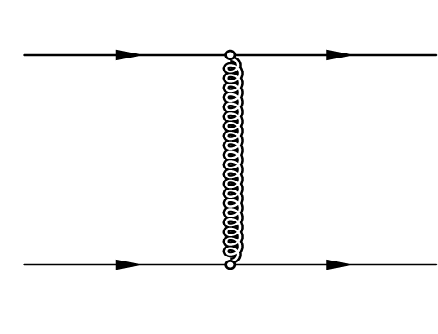}
  \includegraphics[width=.28\textwidth]{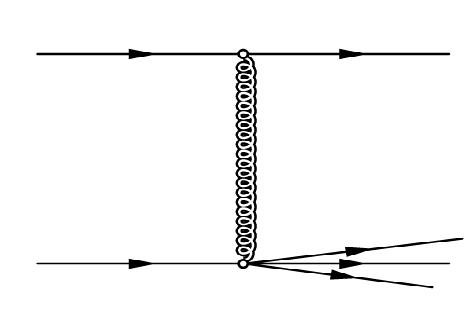}
  \includegraphics[width=.28\textwidth]{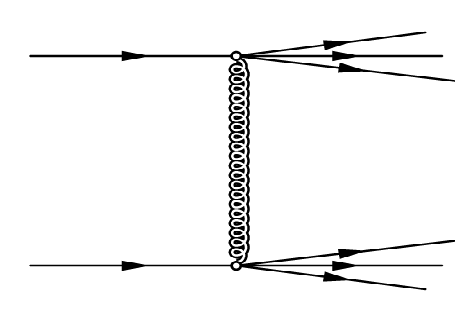}
  \label{fig:topo1}
\end{center}
\end{figure}
\begin{figure}[h]
  \begin{center}
\vspace{-1.6cm}
  \includegraphics[width=.28\textwidth]{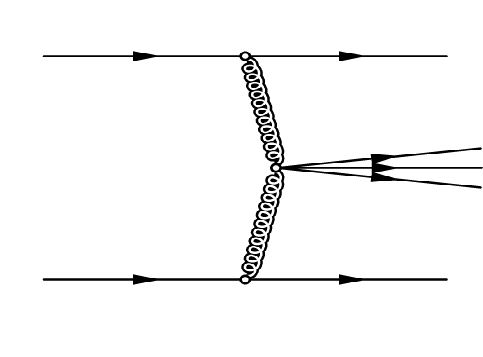}
  \includegraphics[width=.28\textwidth]{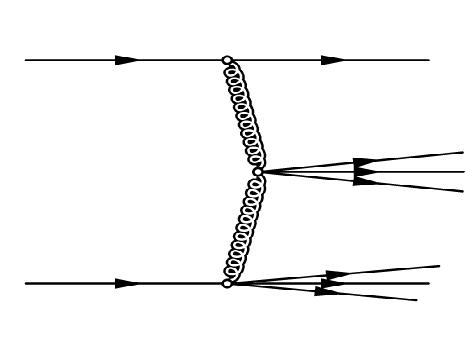}
  \includegraphics[width=.28\textwidth]{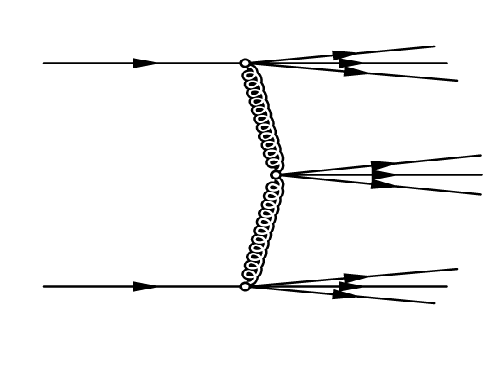}
  \vspace{-.4cm}
  \caption{Top: Single pomeron elastic scatt. (left), and single
    pomeron inelastic scatt. (middle and right).
Bottom: Central exclusive production (left), central production with
proton excitation (middle and right).}
    \label{fig:topo}
\end{center}
\end{figure}
\vspace{-.4cm}
In the top row of Fig. \ref{fig:topo}, the diagrams of strong elastic
and inelastic scattering due to single pomeron exchange are shown. The
corresponding cross sections have been measured at LHC energies by the
TOTEM and the ALICE Collaboration, respectively \cite{TOTEM,ALICE1}.
In the bottom row of Fig. \ref{fig:topo}, the double-pomeron exchange
channel of exclusive production is shown on the left, and central
production with proton excitation in the middle and on the right.
The universality of pomeron exchange can be tested in these central
production channels at high energies, where Regge meson contributions are
negligible as compared to the pomeron exchange.

\section{The spin structure of the soft pomeron}

It is generally agreed that the pomeron exchange carries the vacuum internal
quantum numbers of charge $Q=0$, colour charge $Q_{C}=0$, isospin $I=0$ and
charge conjugation $C=1$. Under intense debate is, however, the spin structure
of the pomeron in the nonperturbative regime of QCD, the soft pomeron
exchange \cite{softPom}. In contrast to the vacuum expectation value of spin
zero, a vector pomeron has been successfully used to describe elastic
proton$-$proton and proton$-$antiproton scattering data, as well as production
of high-$p_{\rm T}$ jets and Drell-Yan pairs \cite{DL1,DL2}. In this vector
approach, the single pomeron amplitudes of proton$-$proton and
proton$-$antiproton elastic scattering carry, however, a relative minus sign.
By use of the optical theorem, this minus sign propagates into an opposite
sign between the proton$-$proton and proton$-$antiproton cross section.

Alternatively, a description of the soft pomeron as an effective \mbox{rank-2}
symmetric tensor has been proposed \cite{Bolz}. Here, the exchanges
of positive and negative charge conjugation are treated as effective tensor
and vector exchanges, respectively, with effective couplings to hadrons
derived from experimental data. The single spin asymmetry data of the STAR
Collaboration have been analysed with the tensor pomeron
model \cite{STAR,Ewerz1}. The results of this analysis disfavour the scalar
pomeron, and are in very good agreement with the tensor pomeron. 
The analysis of central production of scalar and pseudo-scalar mesons
within the vector and tensor pomeron approach finds the azimuthal opening
angle of the two final-state protons to be
a sensitive probe for the two approaches \cite{Lebie1}.

The feasibility of measuring exclusive central production of pseudoscalar
mesons $\eta,\eta'$ at the LHC is of high interest due to the open issues
presented above. The observation of this reaction channel excludes a
scalar pomeron \cite{Nacht}. 

\section{Exclusive eta, eta' production}

Exclusive $\eta,\eta'$-production in proton-proton collisions at the LHC
is a $2 \rightarrow 3$ process,
\begin{eqnarray}
  pp \rightarrow p_{A}p_{B} \;\; \eta(548),\eta'(958). \nonumber
\end{eqnarray}
The three-particle states $p_{A}p_{B}\eta$ and  $p_{A}p_{B}\eta'$ occupy a
five-dimensional phase space. In this study, the five independent parameters
to characterize
this phase space  are the 4-momentum transfers $t_A$ and $t_B$ to the two
protons, the azimuthal angles $\phi_A$ and $\phi_B$ of the two protons,
and the rapidity of the produced meson. We follow the results of
Ref. \cite{Lebie2} for these five phase-space parameters.
\vspace{-3.4mm}
\begin{figure}[h]
    \hspace{6.mm}
  \begin{overpic}[width=.44\textwidth]{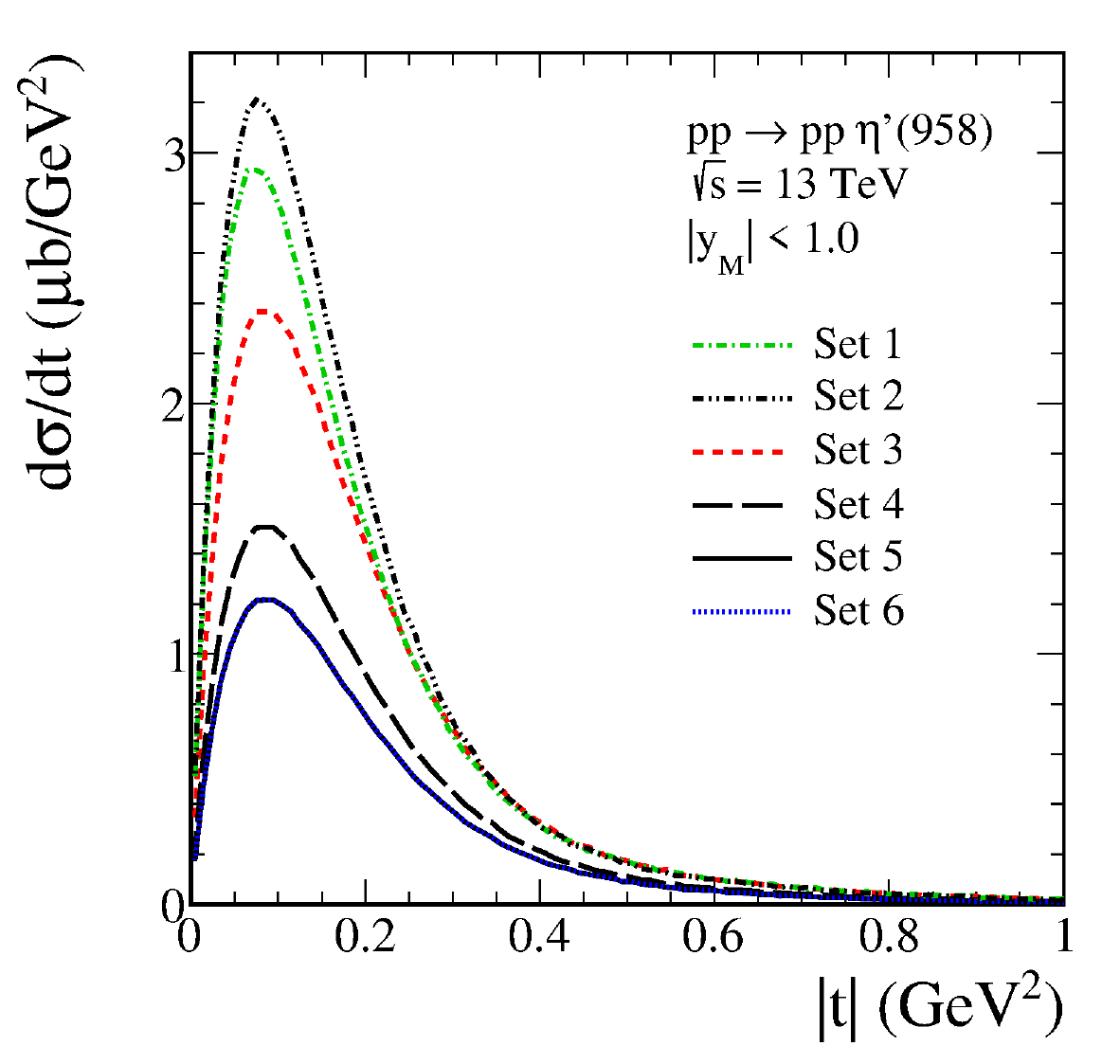}
  \end{overpic}
  \hspace{4.mm}
  \vspace{-4.mm}
  \begin{overpic}[width=.46\textwidth]{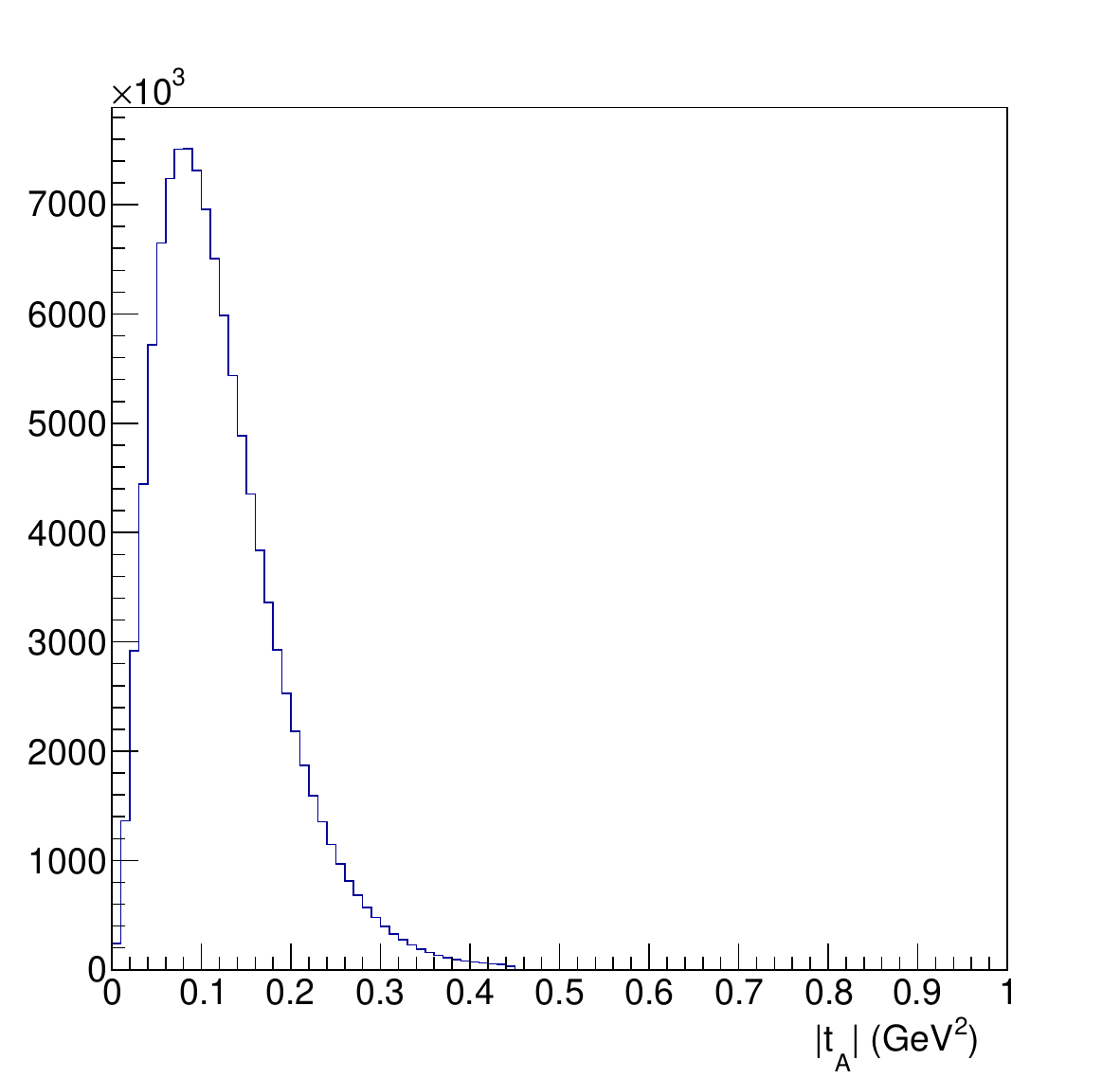}
  \put(-6.,136.){\makebox(0,0){\rotatebox{90.}{\small Number of events }}}
  \end{overpic}
  \vspace{2.mm}
  \caption{Differential cross sections d$\sigma$/dt from Ref. \cite{Lebie2}
    (left), dN/dt generated for this study (right).}
    \label{fig:dsigdt}
\end{figure}
\vspace{-0.3cm}

In Fig. \ref{fig:dsigdt} on the left, the differential cross sections
$d\sigma/dt$ derived for different parameter sets are shown from Ref.
\cite{Lebie2}. On the right, the distribution dN/dt used in this study
is presented.

\begin{figure}[h]
    \hspace{6.mm}
  \begin{overpic}[width=.44\textwidth]{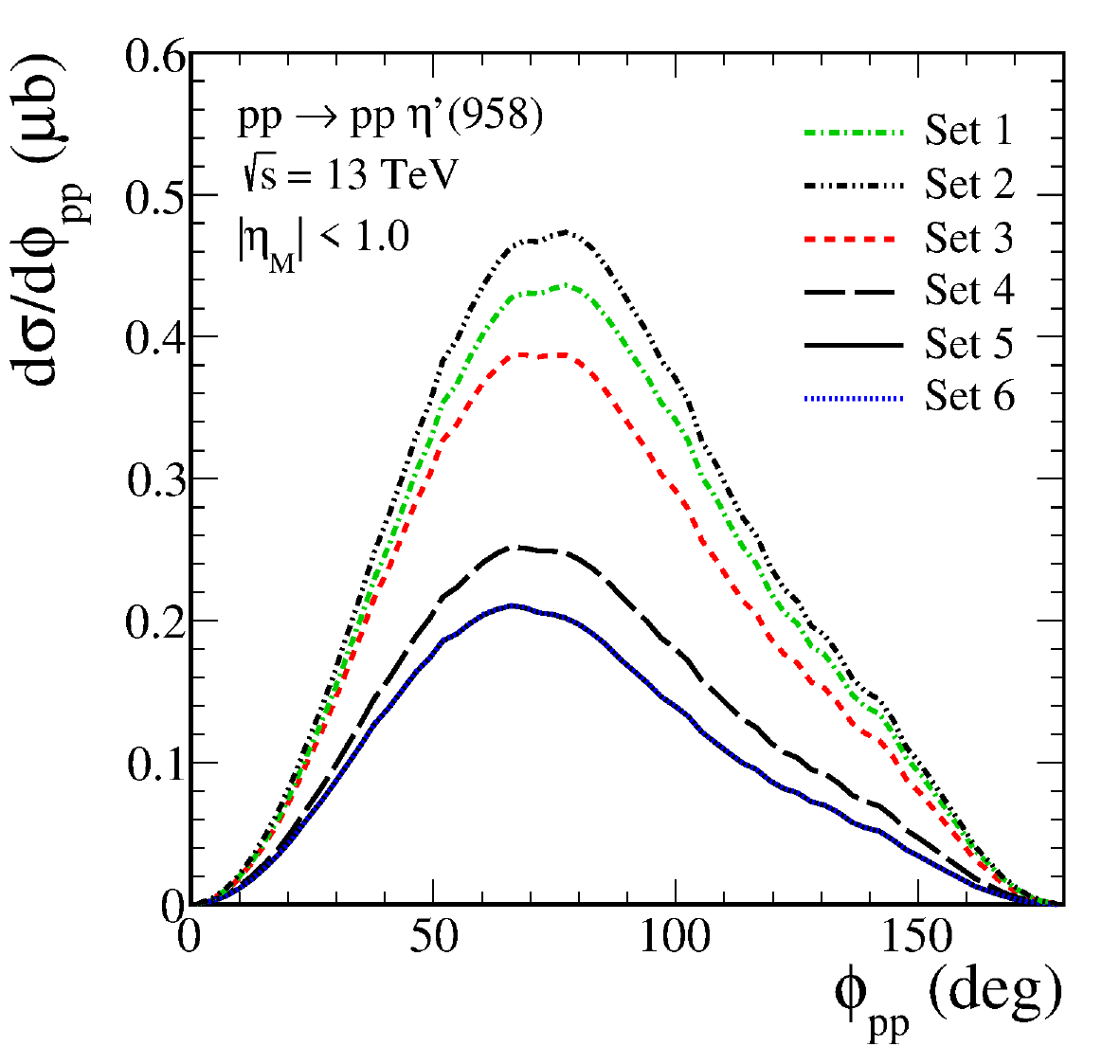}
  \end{overpic}
  \hspace{4.mm}
  \vspace{-4.mm}
  \begin{overpic}[width=.46\textwidth]{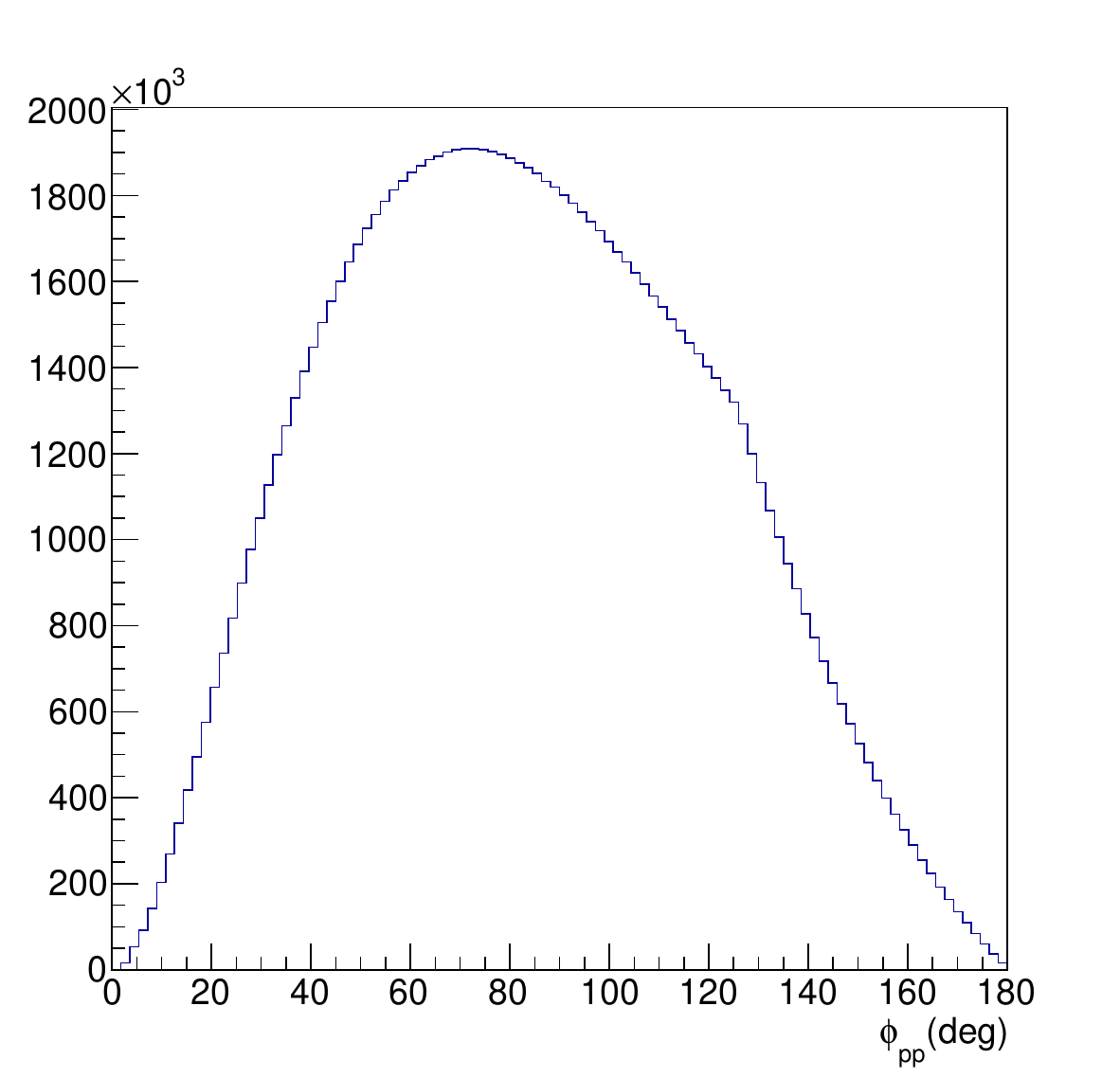}
  \put(-6.,136.){\makebox(0,0){\rotatebox{90.}{\small Number of events }}}
  \end{overpic}
  \vspace{2.mm}
  \caption{Differential cross sections d$\sigma$/d$\phi_{pp}$ from Ref.
    \cite{Lebie2} (left), dN/d$\phi_{pp}$ generated for this study (right).}
    \label{fig:dsigdphi}
\end{figure}
\vspace{-0.3cm}

In Fig. \ref{fig:dsigdphi} on the left, the differential cross sections
$d\sigma/d\phi_{pp}$ derived for different parameter sets are shown from
Ref. \cite{Lebie2}. Here, the variable $\phi_{pp}$ denotes the difference
in the azimuthal angles of the two scattered protons. On the right,
the distribution dN/d$\phi_{pp}$ used in this study is presented.

The rapidity distribution of the $\eta$ and $\eta'$-meson presented
in Ref. \cite{Lebie2} shows a very weak rapidity dependence of a few percent
within the range $-4 < y < 4$. For the study presented here, the rapidity
distributions of the $\eta$ and $\eta'$ are hence taken to be constant.

The final state of the exclusively produced $\eta,\eta'$-meson is composed of
three particles,  the two forward scattered protons and the $\eta,\eta'$-meson.

\subsection{Acceptance of forward protons}

The assessment of acceptance of very forward scattered protons needs a detailed
knowledge of the beam optics parameters. A corresponding calculation
results in the conclusion that \mbox{$\beta^{*}$ = 20 m} and 30 m optics
are feasible at IP2 of the LHC, while not interfering with the luminosity
requirements at the other LHC interaction points \cite{Pascal}. The results
shown below are based on the $\beta^{*}$ = 30 m optics.
The beam spot size at the interaction point in the transverse directions $x,y$
is derived from
\begin{eqnarray}
x_0,y_0 = \sqrt{\epsilon \cdot\beta^{*}} \nonumber
\end{eqnarray}
with $\epsilon$ the beam emittance. The finite divergence of the incoming
beams is neglected.

From the beam optics data file, the parameters are extracted which are
needed to generate the transfer matrices to the positions
of the two forward detectors located 80 m and 112 m away from IP2. 
In order to be detected, the forward scattered protons need to have a
distance larger than 14 $\sigma_{x,y}$ in both transverse directions x and y
in both forward detectors,
with $\sigma_{x,y}$ the beam size in x,y at the detector location.
From the position information of the two forward proton detectors, the
kinematics of the protons at the interaction point is reconstructed.
This reconstruction is presently limited to the transverse kinematics, with
the reconstruction of the longitudinal proton kinematics still under
investigation. The resolution in the position measurement of the forward
scattered protons is taken to be 100 $\mu$m in both transverse directions.

\subsection{Measurement of eta'-meson}

The $\eta'$-meson has different decay channels, in the following this study
is limited to the decay $\eta' \rightarrow  \eta\pi^{+}\pi^{-}$ with a
branching fraction of 42 \% \cite{PDG}.
The matrix elements of this three-body decay have been measured by the BESIII
Collaboration \cite{BESIII}.

\begin{figure}[h]
  \vspace{-2.0mm}
  \begin{overpic}[width=.52\textwidth]{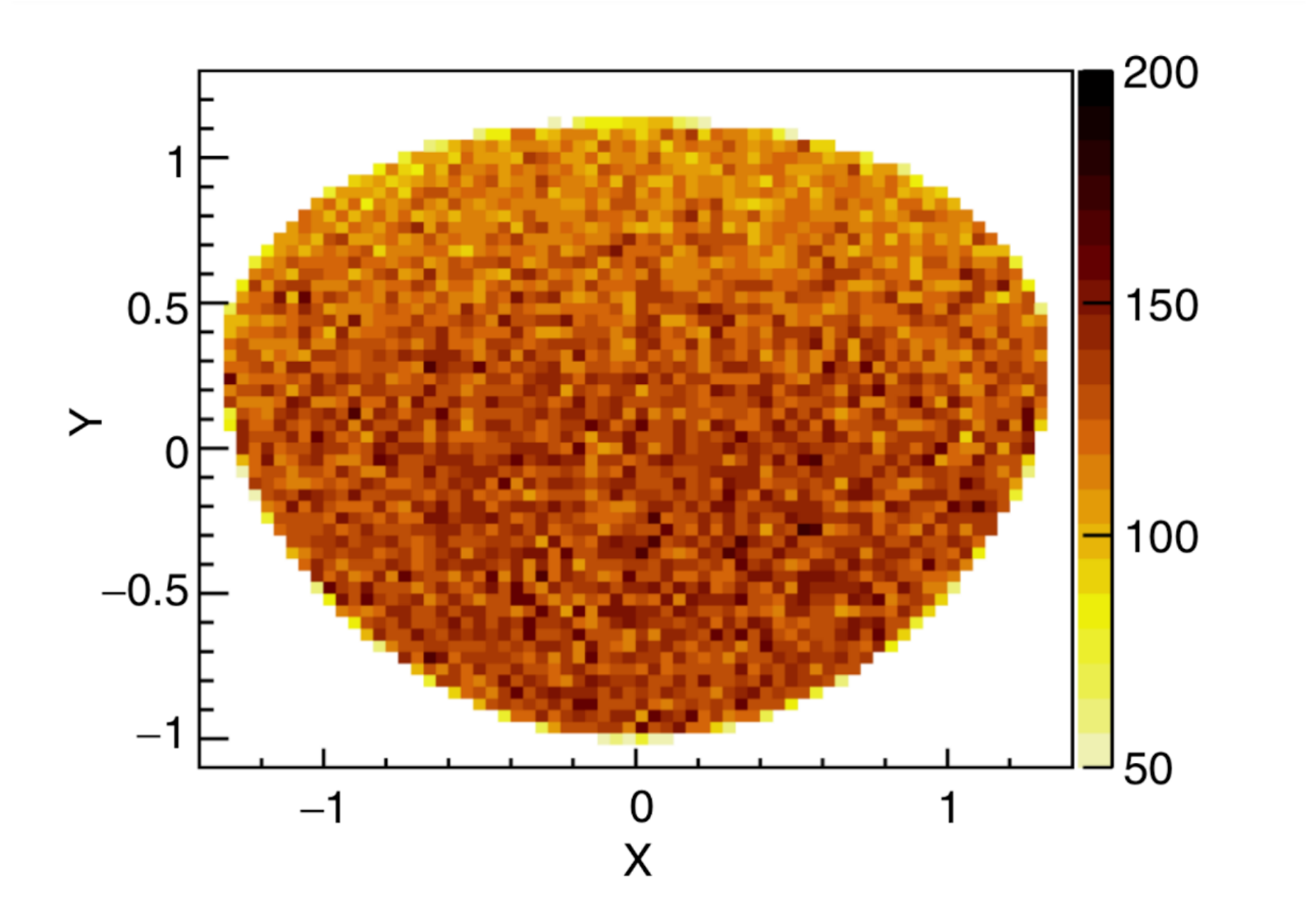}
  \end{overpic}
  \hspace{2.mm}
  \begin{overpic}[width=.44\textwidth]{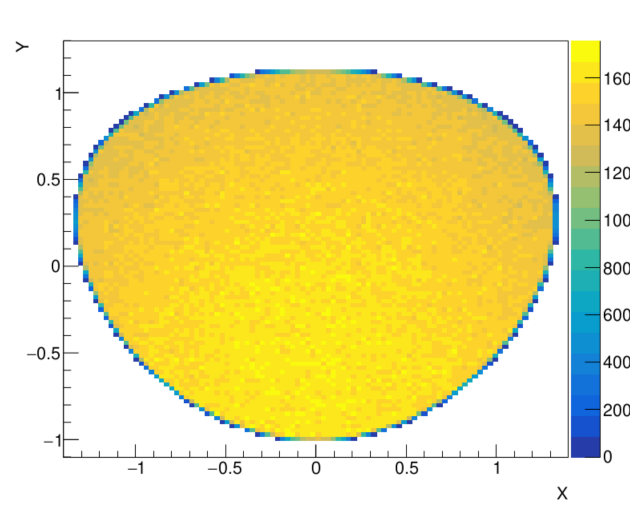}
  \end{overpic}
  \vspace{-.4cm}
  \caption{Dalitz decay distribution $\eta' \rightarrow  \eta\pi^{+}\pi^{-}$
    measured  (left), and generated for this study (right).}
    \label{fig:etap-dal}
\end{figure}

In Fig. \ref{fig:etap-dal}, the measured and generated distribution
of the Dalitz decay $\eta' \rightarrow  \eta\pi^{+}\pi^{-}$ are shown
as function of two kinematical variables $X$ and $Y$, with
$X=\frac{\sqrt{3}(T_{\pi^+}-T_{\pi^-})}{Q}$,
$Y=\frac{m_{\eta}+2.m_{\pi}}{m_{\pi}}\frac{T_{\eta}}{Q}-1$ \cite{BESIII}.

The $\eta$ produced in the $\eta'$-decay is identified in the
2$\gamma$ decay channel with a branching of 40 \%.
\begin{eqnarray}
  pp \rightarrow p_{A}p_{B}  \eta'(958)  \rightarrow p_{A}p_{B} \eta\pi^{+}\pi^{-}
  \rightarrow p_{A}p_{B} \gamma\gamma\pi^{+}\pi^{-} \nonumber
\end{eqnarray}

This final state chosen for the identification of the exclusively produced
$\eta'$-meson necessitates the measurement of six particles:
The two scattered protons, two photons and two oppositely charged pions.

The two charged pions are
restricted to be within the rapidity interval $-1.6 < y < 1.6$, and are 
measured with a  relative momentum resolution of $\frac{\Delta P}{P}$ = 3 \%.
The two photons are restricted to the same rapidity range,
need to have an energy larger than 100 MeV, and are
measured with a relative energy resolution of $\frac{\Delta E}{E}$ = 3 \%.

The final state $\gamma\gamma\pi^{+}\pi^{-}$ can, however, also be produced
by exclusive production and decay of the $f_{1}(1285)$ resonance with
a branching ratio $f_{1} \rightarrow \eta \pi^{+} \pi^{-}$ of 35\%.
\begin{eqnarray}
  pp \rightarrow p_{A}p_{B}  f_{1}(1285)  \rightarrow p_{A}p_{B} \eta\pi^{+}\pi^{-}
  \rightarrow p_{A}p_{B} \gamma\gamma\pi^{+}\pi^{-} \nonumber
\end{eqnarray}

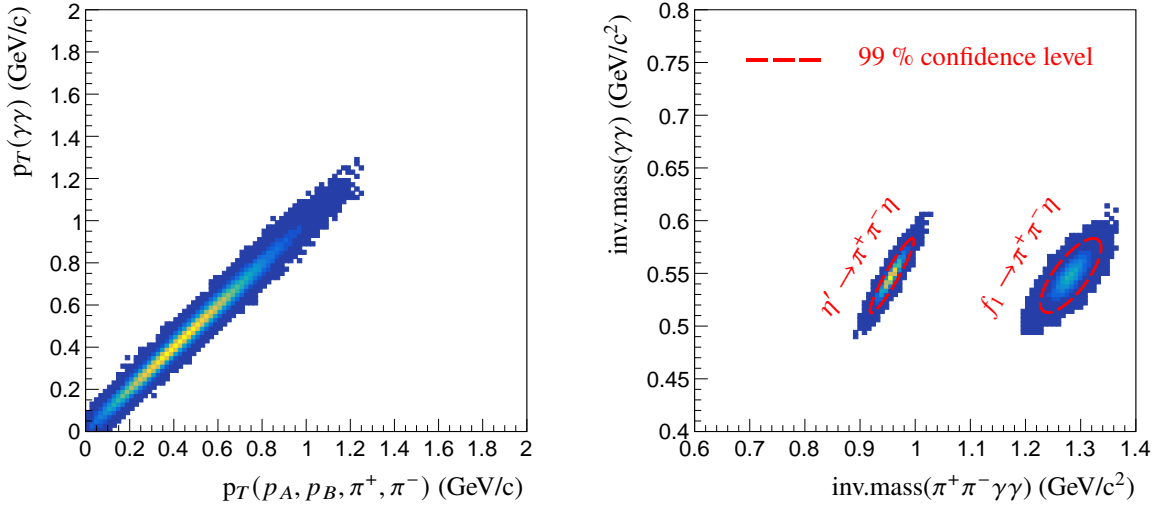
\begin{figure}[h]
  \begin{overpic}[width=.48\textwidth]{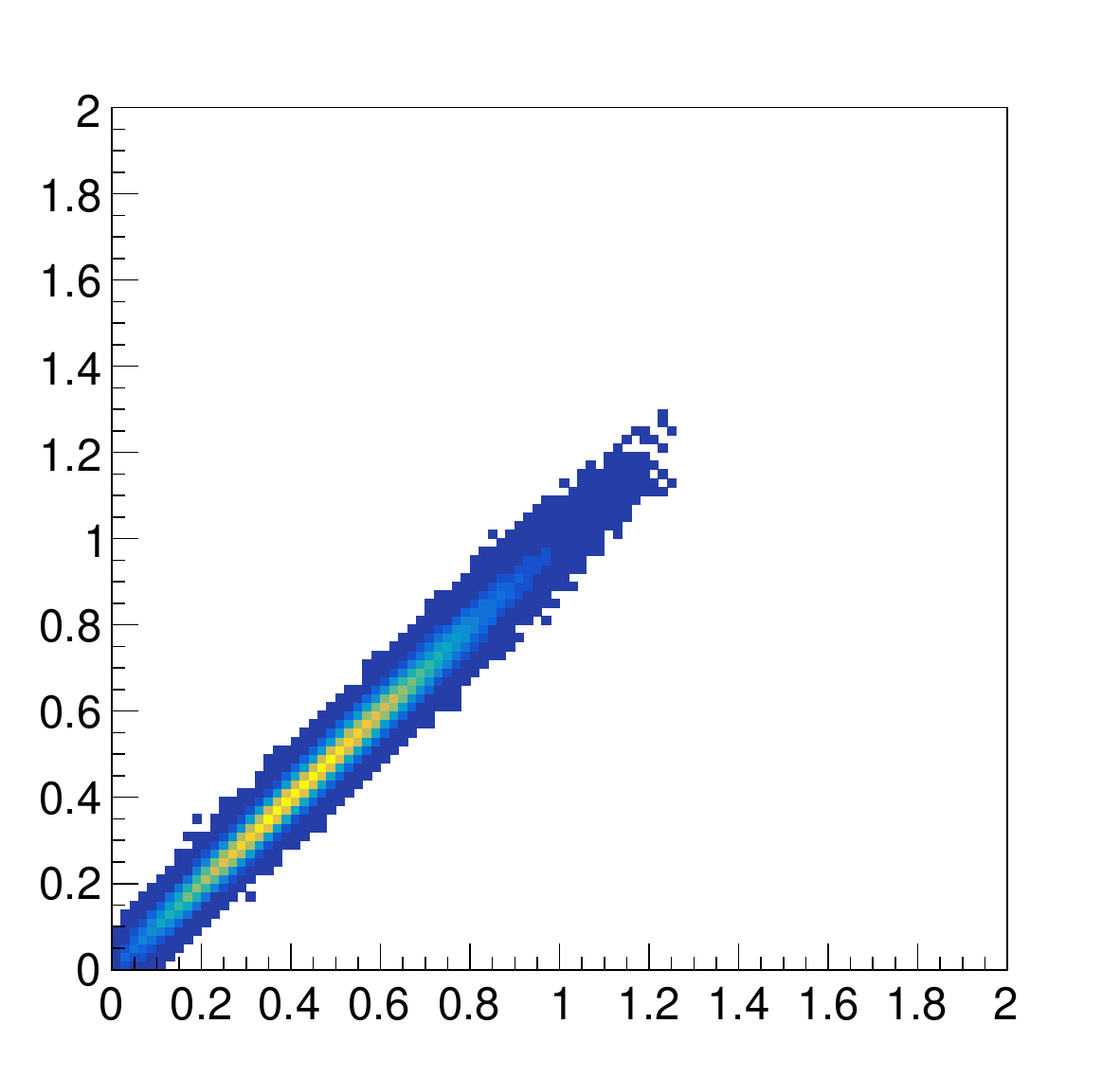}
  \put(72.,-4.0){\small p$_T(p_{A},p_{B},\pi^{+},\pi^{-})$ (GeV/c)}
  \put(-4.,148.){\makebox(0,0){\rotatebox{90.}{\small p$_T(\gamma\gamma)$ (GeV/c)}}}
\end{overpic}
\hspace{.6cm}
\begin{overpic}[width=.48\textwidth]{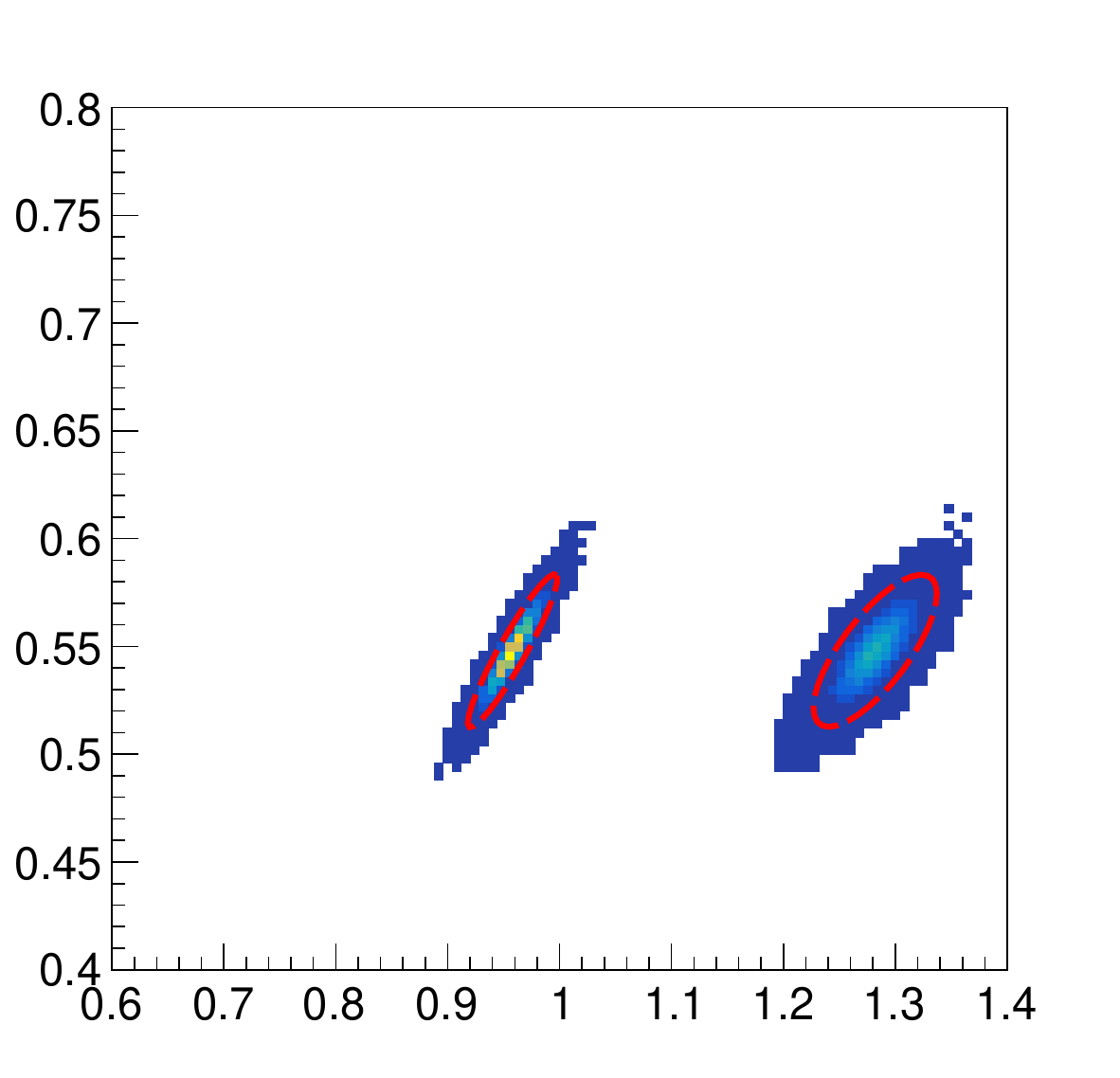}
  \put(72.,-4.0){\small inv.mass($\pi^{+}\pi^{-}\gamma\gamma$) (GeV/c$^{2}$)}
  \put(-8.,132.){\makebox(0,0){\rotatebox{90.}{\small inv.mass($\gamma\gamma$) (GeV/c$^{2}$)}}}
  \put(82.,86.){\makebox(0,0){\rotatebox{60.}{\textcolor{red}{\small $\eta' \rightarrow\!\pi^{+}\pi^{-}\eta$}}}}
  \put(142.,86.){\makebox(0,0){\rotatebox{60.}{\textcolor{red}{\small $f_{1} \rightarrow\!\pi^{+}\pi^{-}\eta$}}}}
\linethickness{0.4mm}
\put(40.,160.){\textcolor{red}{\line(1,0){8}}}
\put(50.,160.){\textcolor{red}{\line(1,0){8}}}
\put(60.,160.){\textcolor{red}{\line(1,0){8}}}
  \put(82.,158.){\textcolor{red}{\small 99 \% confidence level}}
\end{overpic}
  \vspace{-.2cm}
  \caption{Transv. mom. p$_{T}$ of the system ($p_{A},p_{B},\pi^{+},\pi^{-}$)
    vs. trans. mom. p$_{T}$ of the two photons (left),
    inv. mass of the system ($\gamma\gamma\pi^{+}\pi^{-}$)
    vs. inv. mass of the two photons (right).}
    \label{fig:ptptetap}
\end{figure}

In Fig. \ref{fig:ptptetap} on the left, the correlation between the
transverse momenta of the photon pair and the system composed of the two
protons and the two pions is shown. The shown correlation reflects transverse
momentum conservation between initial and final state, and can be used to
identify $p_{A}p_{B}\gamma\gamma\pi^{+}\pi^{-}$ background events.
Such background can further be identified since the
$p_{A}p_{B}\pi^{+}\pi^{-}$ and the  $\gamma\gamma$ systems are back-to-back
in transverse direction, hence have an azimuthal opening angle of $\pi$.
On the right side of Fig. \ref{fig:ptptetap}, the invariant mass of
$\gamma\gamma\pi^{+}\pi^{-}$ on the horizontal axis results from the
primarily produced meson, whereas the photon pair invariant mass on the
vertical axis derives from the secondarily produced $\eta$-meson.
With the chosen experimental resolution specified above, the two resonances
$\eta’(958)$ and $f_{1}(1285)$ can clearly be identified in the
final state $\gamma\gamma\pi^{+}\pi^{-}$.

\subsection{Measurement of eta decay}

The $\eta$-meson has different decay channels, in the following this study is
limited to the decay $\eta \rightarrow  \pi^{+}\pi^{-}\pi^{0}$ with a branching
fraction of 23 \% \cite{PDG}.
The $\pi^{0}$ is identified in the 2$\gamma$ decay channel which has a branching
fraction of 98.8 \%.
\begin{eqnarray}
pp \rightarrow p_{A}p_{B}  \eta  \rightarrow p_{A}p_{B} \pi^{+}\pi^{-}\pi^{0} (BR = 23 \%) \rightarrow p_{A}p_{B} \pi^{+}\pi^{-}\gamma\gamma \nonumber
\end{eqnarray}

This final state  chosen for the identification of the exclusively produced
$\eta$-meson needs the measurement of six particles:
The two scattered protons, two oppositely charged pions and two photons.

The acceptance range, the threshold energy and the resolution for the
measurement of both the pions and the photons are as described above for
the $\eta'$-decay.

The final state $\pi^{+}\pi^{-}\pi^{0}$ can, however, also be produced by
exclusive production and decay of the $\rho(770)$, $\omega(782)$, $\eta'(958)$
and $\phi(1020)$ resonances.
The production of the $\rho(770)$ and $\eta'(958)$ are, however, not
considered in the following due to the small branching fractions
to $\pi^{+}\pi^{-}\pi^{0}$ of $10^{-4}$ and 0.4 \%, respectively.
\begin{eqnarray}
  pp \rightarrow  p_{A}p_{B} \omega(782) \rightarrow & p_{A}p_{B} \pi^{+}\pi^{-}\pi^{0} (BR = 89 \%) \rightarrow & p_{A}p_{B} \pi^{+}\pi^{-}\gamma\gamma \nonumber \\
  pp \rightarrow p_{A}p_{B} {\cal\phi}(1020) \rightarrow & p_{A}p_{B} \pi^{+}\pi^{-}\pi^{0} (BR = 15 \%) \rightarrow & p_{A}p_{B} \pi^{+}\pi^{-}\gamma\gamma  \nonumber
\end{eqnarray}
\vspace{-0.8cm}
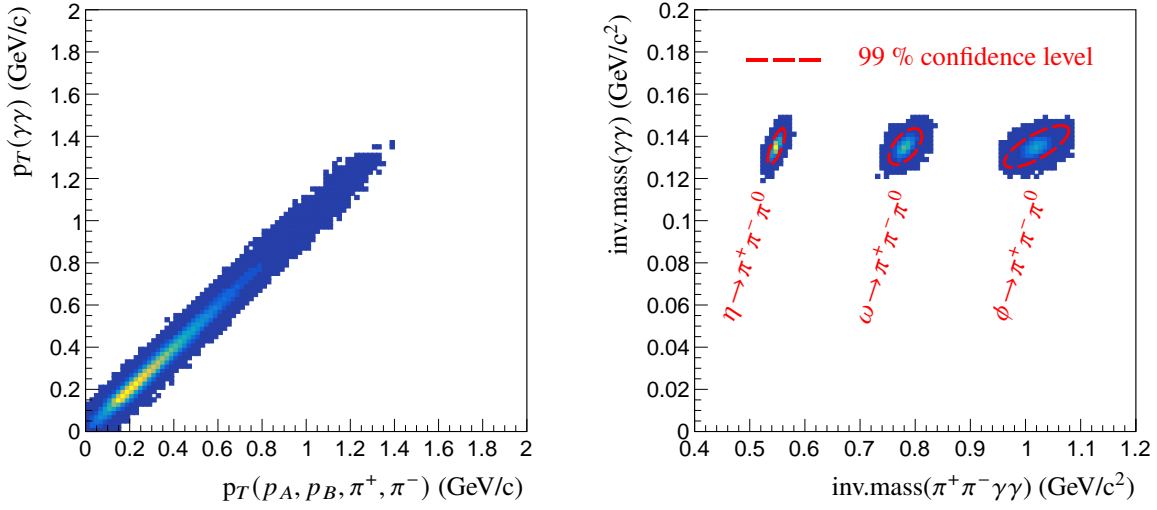
\begin{figure}[h]
\begin{overpic}[width=.48\textwidth]{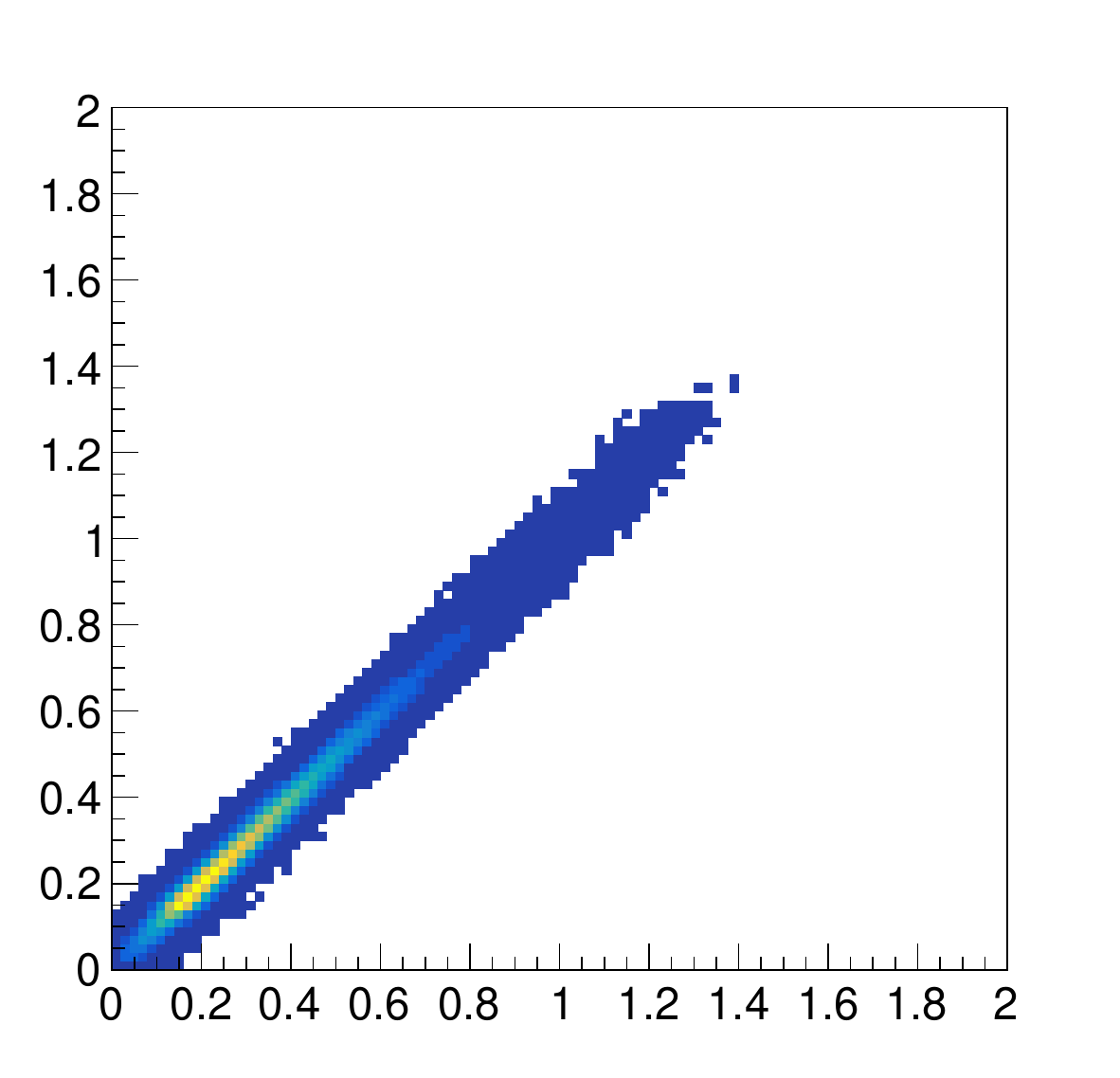}
  \put(72.,-4.0){\small p$_T(p_{A},p_{B},\pi^{+},\pi^{-})$ (GeV/c)}
  \put(-4.,148.){\makebox(0,0){\rotatebox{90.}{\small p$_T(\gamma\gamma)$ (GeV/c)}}}
\end{overpic}
\hspace{0.6cm}
\begin{overpic}[width=.48\textwidth]{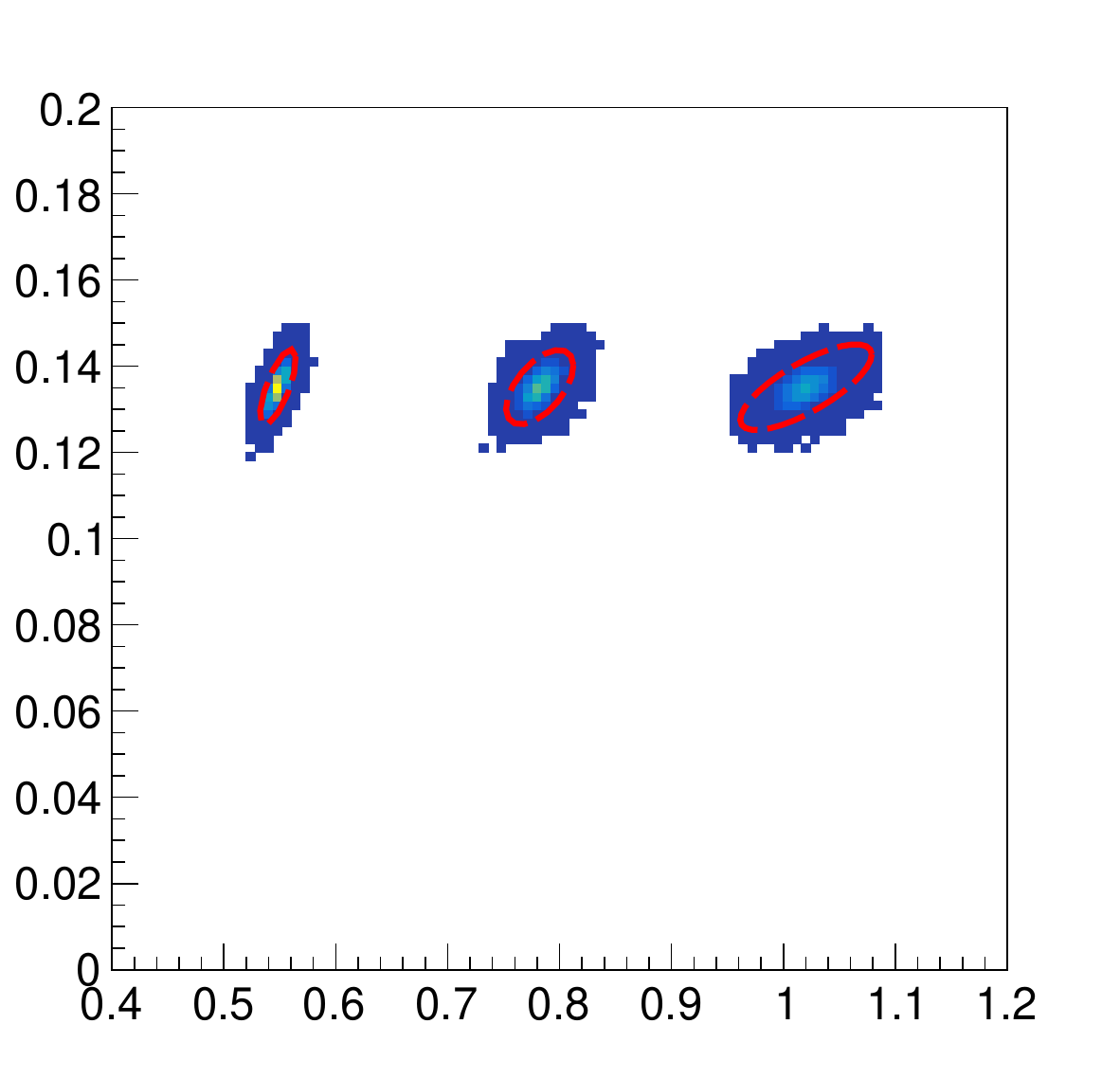}
  \put(72.,-4.0){\small inv.mass($\pi^{+}\pi^{-}\gamma\gamma$) (GeV/c$^{2}$)}
  \put(-8.,132.){\makebox(0,0){\rotatebox{90.}{\small inv.mass($\gamma\gamma$) (GeV/c$^{2}$)}}}
  \put(40.,86.){\makebox(0,0){\rotatebox{72.}{\textcolor{red}{\small $\eta\!\rightarrow\!\pi^{+}\pi^{-}\pi^{0}$}}}}
  \put(90.,86.){\makebox(0,0){\rotatebox{72.}{\textcolor{red}{\small $\omega\!\rightarrow\!\pi^{+}\pi^{-}\pi^{0}$}}}}
  \put(142.0,86.0){\makebox(0,0){\rotatebox{72.}{\textcolor{red}{\small $\phi\!\rightarrow\!\pi^{+}\pi^{-}\pi^{0}$}}}}
\linethickness{0.4mm}
\put(40.,160.){\textcolor{red}{\line(1,0){8}}}
\put(50.,160.){\textcolor{red}{\line(1,0){8}}}
\put(60.,160.){\textcolor{red}{\line(1,0){8}}}
  \put(82.,158.){\textcolor{red}{\small 99 \% confidence level}}
\end{overpic}
  \vspace{-.4cm}
  \caption{Transv. mom. p$_{T}$ of the system ($p_{A}p_{B}\pi^{+}\pi^{-}$)
    vs. trans. mom. p$_{T}$ of the two photons (left),
    inv. mass of the system ($\gamma\gamma,\pi^{+},\pi^{-}$)
    vs. inv. mass of the two photons (right).}
    \label{fig:ptpteta}
\end{figure}

In Fig. \ref{fig:ptpteta} on the left, the correlation between the
transverse momenta of the photon pair and the system composed of the
two protons and the two pions is shown. The shown correlation reflects
transverse momentum conservation between initial and final state, and can
be used to eliminate $p_{A}p_{B}\gamma\gamma\pi^{+}\pi^{-}$ background events.
Such background events can further be identified since the
$p_{A}p_{B}\pi^{+}\pi^{-}$ and the  $\gamma\gamma$ systems are back-to-back
in transverse direction, hence have an azimuthal opening angle of $\pi$.
On the right side of Fig. \ref{fig:ptpteta}, the invariant mass of
$\gamma\gamma\pi^{+}\pi^{-}$ on the horizontal axis results from the mass of
the primarily produced meson, whereas the photon pair invariant mass on the
vertical axis derives from the mass of the secondarily produced $\pi^{0}$-meson.
With the chosen experimental resolution specified above, the resonances
$\eta(548)$, $\omega(782)$ and $\phi(1020)$ can clearly be identified
in the final state $\gamma\gamma\pi^{+}\pi^{-}$.

\section{Summary and outlook}

The prospects for measuring exclusive $\eta,\eta'$-meson
production at the LHC are discussed. Such measurements require
the existence of detectors for measuring the very forward
scattered protons, and detectors at midrapidity for measuring
charged pions, and photons with energies down to 100 MeV.
Results are shown for  identifying the $\eta'$-meson in the
decay $\eta' \rightarrow \gamma\gamma\pi^{+}\pi^{-}$, and for
separating the $f_1(1285)$ which can decay to the same final state.
For the measurement of exclusively produced $\eta$-mesons,
results are shown for the identification of the decay channel
$\eta \rightarrow \gamma\gamma\pi^{+}\pi^{-}$, and for
separating the $\omega(782)$ and $\phi(1020)$ mesons which
can decay to the same final state.

The authors of Ref. \cite{Lebie2} present a range of cross sections
for different parameter sets. As a next step, the achievable
data statistics for exclusive $\eta,\eta'$-production can be
evaluated based on the approach presented here.

\section{Acknowledgements}

The author would like to thank Prof. Nachtmann for many inspiring discussions
on the topics presented here.This work is supported by the German Federal
Ministry of Education and Research under reference 05P24VHA.

\end{document}